\documentclass[aps,prd,twocolumn]{revtex4}

\usepackage{amsfonts}
\usepackage{amsthm}
\usepackage{amsbsy}
\usepackage{amssymb}
\usepackage{amsmath}
\usepackage{graphicx}
\usepackage{color}

\begin{document}

\date{\today}


\title{Hawking radiation from ultrashort laser pulse filaments}

\author
{F. Belgiorno,$^{1}$ S.L. Cacciatori,$^{2,3}$ M. Clerici,$^{3}$ V. Gorini,$^{2,3}$\\
G. Ortenzi,$^{4}$ L. Rizzi,$^{3}$ E. Rubino,$^{3}$ V.G. Sala,$^{3}$ D. Faccio$^{3,5\ast}$}

\address{$^{1}$Dipartimento di Fisica, Universit\`a degli Studi di Milano, Via Celoria 16, IT-20133 Milano, Italy\\
$^{2}$INFN sezione di Milano, via Celoria 16, IT-20133 Milano, Italy\\
$^{3}$Dipartimento di Fisica e Matematica, Universit\`a dell'Insubria, Via Valleggio 11, IT-22100 Como, Italy\\
$^{4}$Dipartimento di Matematica e Applicazioni, Universit\`a di Milano-Bicocca, Via Cozzi 53, IT-20125 Milano, Italy\\
$^{5}$School of Engineering and Physical Sciences, David Brewster Building, Heriot-Watt University, Edinburgh, Scotland
EH14 4AS, UK}

\email{ E-mail: d.faccio@hw.ac.uk}



\begin{abstract}
Event horizons of astrophysical black holes and gravitational analogues have been predicted to excite the quantum vacuum and give rise to the emission of quanta, known as Hawking radiation. 
 We experimentally create such a gravitational analogue using ultrashort laser pulse filaments and our measurements 
demonstrate a spontaneous emission of photons that confirms theoretical predictions.\end{abstract}

\maketitle

\begin{figure}[t]
\centering
\includegraphics[width=8cm]{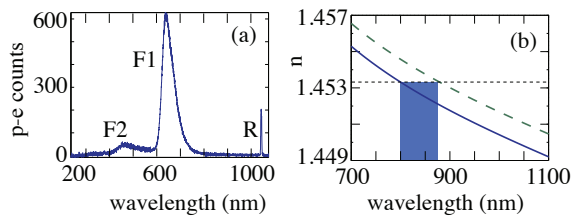}
\caption{\label{fig:prediction} (a) Measured CCD photo-electron (p-e) counts generated by the fused silica fluorescence spectrum. (b) Prediction of the Hawking emission spectral range for the case of fused silica: $n_0$ (blue solid line) and $n_0+\delta n$ (green dashed line) with $\delta n=0.001$.  The shaded area indicates the spectral emission region predicted for a Bessel filament.}
\end{figure}

\begin{figure}[t]
\centering
\includegraphics[width=7cm]{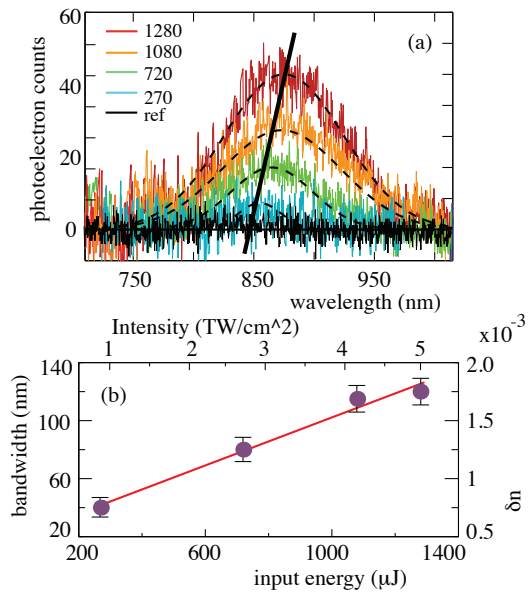}
\caption{\label{fig:results} (a) Spectra generated by a Bessel filament. Five different curves are shown corresponding to a reference spectrum obtained with a Gaussian pulse (black line) and indicated  Bessel energies (in $\mu$J) . Dashed lines are guides for the eye. The solid line connects the spectral peaks, highlighting the $\sim40$ nm shift with increasing energy, in close agreement with the predicted 45 nm shift. (b) Bandwidths at FWHM and RIP $\delta n$  versus input energy and intensity. Solid line: linear fit $\delta n = n_2I$ with $n_2=2.8\times10^{-16}$ cm$^2$/W.}
\end{figure}

\begin{figure}[t]
\centering
\includegraphics[width=8cm]{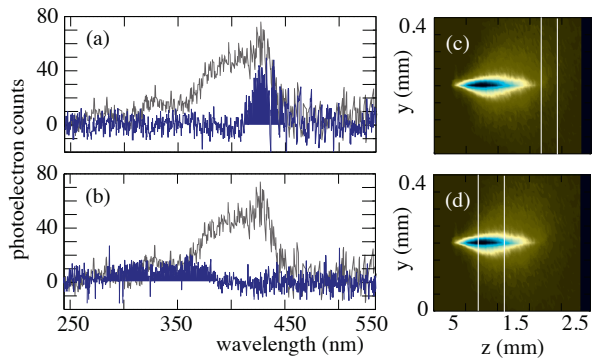}
\caption{\label{fig:filament} Spectra generated by the spontaneous filament. (a)-(b) shaded areas: spectra measured for two different positions of the imaging spectrometer input slit. The light gray curves in both figures show the spectrum measured with the input split fully open. The filament, imaged from the side at 90 deg., is shown in (c) and (d). The vertical white lines show the position of the input slit.}
\end{figure}

In 1974 Hawking predicted that the space-time curvature at the event horizon of a black hole is sufficient to excite photons out of the vacuum and induce a continuous flux, referred to as Hawking radiation \cite{hawking1,hawking}. In a simplified description of the process, vacuum fluctuation pairs close to the horizon are split so that the inner photon falls in and the outside photon escapes away from the black hole. As the outgoing photon cannot return to the vacuum, it necessarily becomes a real entity, gaining energy at the expense of the black hole.  It was soon realized that the essential ingredient of Hawking radiation was not the astrophysical black hole itself but rather the space-time curvature associated to the event horizon \cite{unruh,barcelo_review,visser,unruh2,visser2}. There are a wealth of physical systems that may exhibit event horizons ranging from flowing water or Bose-Einstein-Condensates to a moving refractive index perturbation (RIP) in a dielectric medium \cite{unruh,novello_book,barcelo_review,philbin,faccio}. In a few words, referring to the case of optical pulses in a dielectric medium proposed by Philbin et al. \cite{philbin}, a laser pulse with large intensity, $I$, propagating in a nonlinear Kerr medium will excite a RIP given by $\delta n=n_2I$ where $n_2$ is the so-called nonlinear Kerr index \cite{boyd}. Light experiences an increase in the local refractive index as it approaches the RIP and is thus slowed down. By choosing appropriate conditions (frequency of the light and velocity of the RIP)  it is possible to bring the light waves to a standstill in the reference frame comoving with the RIP,  thus forming a so-called white hole event horizon, i.e. a point beyond which light is unable to penetrate. A similar mechanism may be observed with water waves or with any kind of waves in a flowing medium, so that the formation of an analogue event horizon  is a rather universal phenomenon that may be studied in accessible laboratory conditions. What remains to be established, is whether Hawking radiation is actually emitted in the presence of an event horizon of any kind, be it analogue or astrophysical.\\
Recently, an alternative approach was proposed for generating
controllable RIPs, namely ultrashort laser pulse filamentation \cite{faccio}.\\
Ultrashort laser pulse filaments are intense laser pulses in a transparent Kerr medium (i.e. with a third 
order optical nonlinearity) characterized by a high-intensity spike that propagates apparently without diffraction over distances much longer 
than the Rayleigh length associated to the spike dimensions and have been proposed for many applications ranging from white light generation to the control of atmospheric conditions \cite{couairon,berge,science1,science2,KaspNature}. Filaments may either occur spontaneously when a powerful Gaussian shaped beam is 
loosely focused into the Kerr medium \cite{couairon,berge} or, alternatively, they may be induced by pre-shaping the laser pulse into a Bessel 
beam \cite{pole1,pole2}.\\
In this Letter we use ultrashort laser pulse filaments to create a traveling RIP in a transparent dielectric medium (fused silica glass) and we report experimental evidence of photon emission that on the one hand bears the characteristics of Hawking radiation and on the other is distinguishable and thus separate from other known photon emission mechanisms. We therefore interpret the observed photon emission as an indication of Hawking radiation induced by the analogue event horizon.\\
Neglecting any dependence on the transverse coordinates and dispersion it is possible to give a complete description of the event horizon associated to the RIP and calculate a blackbody temperature of the emitted photons in the laboratory reference frame \cite{philbin,belgiorno}. However, we point out that the dielectric medium in which the RIP is created will always be dominated by optical dispersion and this in turn implies that in real settings the spectrum will not be that of a perfect blackbody and that in any case only a limited spectral portion of the full  spectrum will be observable. To show this last point we describe the RIP  as a perturbation induced by the laser pulse on top of a uniform, dispersive background refractive index $n_0$, i.e. $n(z,t,\omega)=n_0(\omega)+\delta n f(z-vt)$, where $\omega$ is the optical frequency, $f(z-vt)$ is a function bounded by 0 and 1, 
that describes the shape of the laser pulse. 
 In the reference frame co-moving at velocity $v$ with the RIP, the event horizon in a 2D geometry is defined by  $c/v=n$ which admits solutions only for RIP velocities satisfying the inequality \cite{belgiorno}
\begin{equation}\label{horizon}
\frac{1}{n_0(\omega)+\delta n}<\frac{v}{c}<\frac{1}{n_0(\omega)}
\end{equation}
Therefore, Eq.~(\ref{horizon}) predicts an emission spectrum with well-defined boundaries and it is precisely this feature of the spectral 
emission that we consider to be peculiar to analogue Hawking radiation in the present settings.\\
\begin{figure}[!t]
\centering
\includegraphics[width=7cm]{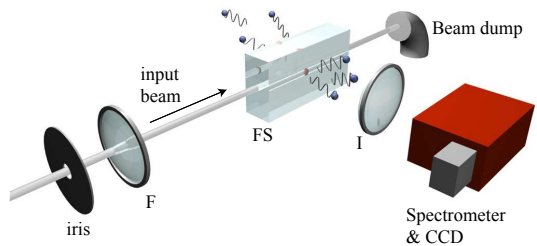}
\caption{\label{fig:layout} Experimental layout used for detecting analogue Hawking radiation. The input laser pulse is focused into a sample of fused silica (FS) using an axicon or lens (F).  An imaging lens (I) collects the photons emitted at 90 deg and sends them to an imaging spectrometer coupled to a cooled CCD camera. }
\end{figure}
We note that two horizons may be associated to each RIP: the leading edge is the analogue of a black hole horizon whilst the trailing edge has time reversed features and is the analogue of a white hole horizon \cite{philbin,faccio}. \\
The experimental layout is shown in Fig.~\ref{fig:layout}. The laser pulses are provided by a regeneratively amplified, 10 Hz repetition rate 
Nd:glass laser. The pulse duration is 1 ps and maximum energy is 6 mJ. 
The Bessel pulse filament with a cone angle $\theta=7$ deg is generated by a 20 degree fused silica axicon (conical lens), F, placed directly in 
front of the 2 cm fused silica Kerr sample where the RIP is generated. The input energy is varied in the 100-1200 $\mu$J range.\\
Radiation from the filament is then collected at 90 degrees with respect to the laser pulse propagation axis using a lens that images the filament on to the input slit of an imaging spectrometer. The spectrum is then recorded with a 16 bit, cooled CCD camera. This arrangement, in particular detection at 90 degrees, was chosen in order to strongly suppress or eliminate any spurious effects. More specifically:\\
(i) Cerenkov-like radiation, i.e. in this context, radiation from a superluminal perturbation: spontaneous emission from a strictly superluminal perturbation has been considered in detail in Ref.~\cite{belgiorno_superluminal}. The most relevant difference with respect to the present measurements is that Cerenkov-like emission occurs with no upper bound in the spectral emission window, in contrast to the limited Hawing spectrum given by Eq.~(\ref{horizon}).\\
(ii) Four wave mixing (FWM) and self phase modulation (SPM) will not occur at 90 degrees due to phase-matching constraints which imply that any newly generated frequencies will be generated at small angles with respect to the propagation axis \cite{gadonas}. In any case we directly verified that for the relatively large $\theta=7$ deg Bessel cone angle used in our experiments, no FWM, SPM or spectral broadening was observed at any angle, even in the forward direction.\\
(iii) Rayleigh scattering will occur at 90 degrees only for vertically polarized light and the scattering process will maintain the polarization state. In our experiments we used horizontally polarized light and in any case, in virtue of point (i), there is no input or generated light at the frequencies relevant for the present experiment.\\
(iv) Fluorescence is certainly the main problem in these experiments. It bears many features in common with Hawking radiation yet, it may still be clearly distinguished from the latter. Figure~\ref{fig:prediction}(a) shows the overall spectrum measured at 90 degrees and integrated over 30 laser pulse shots. In the figure, R, F1 and F2 indicate the laser pulse induced spontaneous Raman, nonbridging oxygen hole centre (NBOHC) and oxygen deficient centre (ODC) fluorescences, respectively. These fluorescence peaks are well documented features of fused silica \cite{fluoro1,fluoro2}. This therefore  allows to fit the peaks with Gaussian functions (in the frequency domain) and subsequently subtract out the fluorescence signals, thus leading to greatly improved contrast and cleaner spectra. \\
Nevertheless, looking at the fluorescence spectrum it is clear that, given the possibility to tune the Hawking emission window by tuning the group velocity of the laser pulse,  the most advantageous situation is that in which the window is located between 800 nm and 900 nm where there is no known fluorescence emission and moreover the CCD  response is maximum. In fused silica, $n_2 \sim 3\cdot10^{-16}$ cm$^2$/W  and typical intensities of $I\sim10^{12}-10^{13}$ W/cm$^2$ may be obtained, thus leading to 
a $\delta n\sim 10^{-2}-10^{-4}$. Therefore, if we take for example a Gaussian pulse centred at 1055 nm, with $\delta n=0.001$ and group velocity, $v_G=d\omega/dk$, determined solely by material dispersion, then  according to Eq.~(\ref{horizon})  we would expect emission in the region between 500 nm and 510 nm. This would clearly fall in between the F1 and F2 fluorescence peaks and leads to noisy results even after subtraction of the fluorescence signal. 
By using Bessel pulses we were able to solve this problem. Figure~\ref{fig:prediction}(b) shows the  Hawking spectral window predicted from Eq.~(\ref{horizon}), where the dashed line shows the value $c/v_B$ with $\theta=7$ deg and the Bessel pulse velocity is $v_B=v_G/\cos\theta$. This line intersects the refractive index curves $n$ and $n+\delta n$ with $\delta n=0.001$ delimiting a window (shaded area) which now lies in the desired region, i.e. between 800 nm and 875 nm.
Figure~\ref{fig:results}(a) shows the resulting spectra, integrated over 3600 laser shots. The black line shows a reference spectrum obtained 
from a Gaussian pulse, that clearly shows the absence of any signal notwithstanding the same peak intensity of the Bessel pulses and thus underlining the absence of any fluorescence or other possible emission signals when the RIP $v$ does not 
satisfy Eq.~(\ref{horizon}). The other four curves show the emitted spectra for four different input energies of the Bessel filament, as 
indicated in the figure. A clear photon emission is registered in the  wavelength window  predicted by Eq.~(\ref{horizon}).   Moreover we verified that the emitted radiation was unpolarized (data not shown) thus further supporting the interpretation of a spontaneous emission. 
In addition, the measurements clearly show that the bandwidth of the emission is  increasing with the input energy. The Bessel pulse 
intensity evolution along the propagation direction $z$ may be estimated analytically from the input energy using the formula 
$I(z)=2\pi I_0 k z\tan^2(\theta)\exp[-{(z^2\tan^2\theta)}/{w^2_0}]$ \cite{friberg} 
where $k=\omega/c$ and $I_0$ and $\sqrt{2} w_0$ are the input Gaussian peak intensity and radius at $1/e^2$, respectively. By fitting the measured spectra with Gaussian functions [dashed curves in Fig.~\ref{fig:results}(a)] we may therefore estimate the bandwidth as a function of input energy and Bessel intensity.  Moreover,  using the fused silica dispersion relation shown in Fig.~\ref{fig:prediction}(b) it is possible to map the bandwidth into values of $\delta n$ so that we finally obtain Fig.~\ref{fig:results}(b) that shows the bandwidth and the $\delta n$ as a function of input energy and Bessel pulse peak intensity (at $z=1$ cm where measurements were performed). There is a clear linear dependence which is in qualitative agreement with the fact that, according to our interpretation, the emission bandwidth is predicted to depend on $\delta n$ which in turn is a linear function $\delta n = n_2I$ of the pulse intensity.  
The slope of the linear fit gives 
$n_2 = 2.8\pm0.5\times10^{-16}$ cm$^2$/W which is in good agreement with the tabulated value of $n_2 \sim 3\times10^{-16}$ cm$^2$/W \cite{desalvo,couairon}. Therefore this shows that there is also an agreement at the quantitative level between the measurements and the model based on Hawking-like radiation emission.\\
In Fig.~\ref{fig:filament} we present additional data regarding Hawking emission from a spontaneous filament obtained by replacing the axicon with a 20 cm focal length lens and thus loosely focusing a 50 $\mu$J Gaussian pulse into the fused silica sample. The spontaneous nonlinear Kerr dynamics lead to the formation of a filament \cite{couairon} that is very similar to the Bessel filament in the sense that it is characterized by a very high-intensity peak that propagates over long distances, thus creating a strong RIP in the medium that may  be expected to excite Hawking radiation in a similar fashion to that illustrated above. We underline some important differences with respect to the Bessel filament: the spontaneous  filament  is characterized by an intense pulse that travels slower than the input Gaussian pulse, i.e. $v<v_G$ (see \cite{faccio2} for a detailed description). Moreover the pulse accelerates during propagation so that a certain range of velocity values is covered during propagation, with low velocity at the beginning of the filament and higher velocity at the end. This has the effect of broadening the emission window, which is predicted from Eq.~(\ref{horizon}) to be between 270 nm and 450 nm. The light gray-curve in Fig.~\ref{fig:filament}(a) shows the spectrum measured when collecting light from the whole filament. As can be seen, the  measured spectral bandwidth agrees well with the prediction. We also note that no emission was observed in this region with the Bessel filament, indicating that this is not a fluorescence signal. Moreover, by shifting and closing the input slit of the imaging spectrometer, we could collect spectra of the ending and beginning sections of the filament, that are shown as dark-filled curves in Figs.~\ref{fig:filament}(a) and (b), respectively. Remarkably, different sections of the filament emit only at selected portions of the overall spectrum. We underline that, accounting for the different pulse velocities in the different filament sections, this striking behavior is in quantitative agreement with the predictions of Eq.~(\ref{horizon}).\\
%
Finally, we note that Eq.~(\ref{horizon}) and subsequent analysis, refers to the existence of phase velocity horizons. However, other analogue systems rely on the existence of group velocity horizons, i.e. points at which $v=v_G$ (see e.g. \cite{rousseaux,unruh2010}) so that there is still an open question regarding the relative role of the two different types of horizon. In the specific case of the Bessel pulse used in our experiments, a group horizon does not even exist, i.e. there are no frequencies that satisfy $v_G(\omega)=v_B$. Our measurements therefore indicate that, at least in the RIP setting, the phase velocity horizon alone may lead to photon emission in the predicted spectral window.\\
In summary, the same physics that underlie black hole evaporation in the form of Hawking radiation may be found and studied in other, more accessible systems. Our measurements highlight spontaneous emission of Hawking radiation from an analogue event horizon generated by an  ``evaporating'' refractive index perturbation and
suggest a path towards the experimental study of  phenomena traditionally relegated to the areas of quantum gravity and astrophysics. \\
The authors acknowledge P. Di Trapani and A. Couairon for technical assistance and discussions with I. Carusotto, A. Recati, S.M. Barnett, L. Vanzo, S. Zerbini, M. Kolesik and U. Leonhardt.

\end{document}